\documentclass{aa}  

\usepackage{graphicx}
\usepackage{txfonts}
\usepackage{lipsum}
\usepackage{subcaption}         
\usepackage{lscape}             
\usepackage{placeins}           

\usepackage{xcolor}
                                
\begin{document}


   \title{Complex hydrogen chemical equilibrium and Gaia low mass problem in cool white dwarfs}



   \author{Piotr M. Kowalski}

   \institute{Institute of Energy Technologies - Theory and Computation of Energy Materials (IET-3), Forschungszentrum J\"{u}lich GmbH, 52425 J\"{u}lich, Germany. Correspondence \email{p.kowalski@fz-juelich.de}            
    \and Jülich Aachen Research Alliance
JARA Energy \& Center for Simulation and Data Science (CSD)
52428 Jülich, Germany, }


 
  \abstract
   {Large Gaia data set shows substantial misfit between models and observation for cool white dwarfs with $T_{\rm eff}<6000\,\rm K$, resulting in severe underestimation of masses of these stars.}
   {We aim to understand the underlying modelling issues.} 
   {State of the art atmosphere models have been applied to analyse the Gaia DR3 sample of white dwarfs as well as quantum mechanical calculations to quantify formation and stability of different hydrogen species in the atmospheres of these stars.}
   {We reconcile the models and observations when we artificially suppress formation of $\rm H_3^+$ species, a process which substantially alters the chemical equilibrium at $T_{\rm eff}<6000\,\rm K$, resulting in an overabundance of free electrons and $\rm H^-$, and strengthening of $\rm H^-$ bound-free absorption. Removing the $\rm H_3^+$ species from chemical equilibrium consideration makes ionization of hydrogen atoms the main source of free electrons, with the resulting models reproducing well the Gaia white dwarfs cooling branch. Because $\rm H_3^+$ must form under the considered conditions, likely it is the overestimation of its partition function and resulting abundance or the formation of $\rm H_3^-$ or another anionic species that suppresses the formation of $\rm H^-$ as a countercharge for $\rm H_3^+$ in current models.}
   {Chemical equilibrium in cool, hydrogen atmospheres white dwarfs must be reconsidered in respect to the abundance of $\rm H_3^+$ species and presence of unaccounted charge species.}


   \maketitle
   \nolinenumbers

\section{Introduction}

The large amount of data delivered by Gaia put models of white dwarf stars to the test. Several studies published in the last three decades established large confidence on the predictability of the spectra and photometries of hydrogen atmosphere stars representing the majority of white dwarfs \citep{KS06,KKH09,KKR09,KMW10,GBD12,DKP12,KTK12,SHK14,BDT19,CBB23}. Especially after the introduction of hydrogen Lyman (Ly) $\alpha$ red wing opacity by \citet{KS06}, spectra of the majority of such stars could be well fitted, including the stars at the cooler end (with $T_{\rm eff}\sim4000\,\rm K$, e.g., \citet{KKR09,KMW10,SHK14,DKP12}). Nevertheless, although several thousands of white dwarfs have been successfully identified in Gaia data, allowing the construction of a complete white dwarf Hertzsprung-Russell diagram, all studies up to date show significant divergence between the models and the data in Gaia $G$ vs $G_{BP}$ $-$ $G_{RP}$ cooling track for stars with $T_{\rm eff}$ below $\sim6000\,\rm K $ (e.g., \citet{CBB23,BTK23,STK25}). This discrepancy leads to severe underestimation of masses of cool stars by up to $0.2\rm \, M_{\sun}$ and in inability to perform correct analysis of the cooling sequence in terms of, for instance, age \citep{STK25}.

The majority of white dwarf stars, in particular the cool objects, have hydrogen-dominated atmosphere with some minority of objects showing helium dominated atmospheres \citep{KS06,CBB23}. While the more extreme helium atmosphere stars, with density reaching that of dense fluid (up to a few $\rm g/cm^3$, \citet{K14,SBT22}), represent a challenge to the modelling, as several dense fluid effects have to be accounted for (e.g., \citet{KS04,K06,K14,BKD17}), the hydrogen atmospheres are less extreme and should be well described by the ideal gas physics and chemistry, as indicated by the good performance of models in the above mentioned studies. The detection of a discrepancy between models and Gaia data comes thus as a surprise and a puzzle.

Some recent studies attempted an analysis of the problem, looking for inaccuracies in the description of main absorption mechanisms as a main cause. \citet{BTK23} illustrated that an ad hoc increase of Ly $\alpha$ opacity of \citet{KS06} by a factor of 5 could correct the mass problem, calling for a revision of the model of this absorption mechanism. Although plausible, so severe uncertainty in Ly $\alpha$ absorption would contradict above mentioned successful fits to several cool white dwarf stars. Nevertheless, \citet{STK25} attempted revision of this opacity mechanism, including effects of multiple perturbers and detailed angular dependence analysis of $\rm H_2-He$ collisions. Although the resulting models lead to some minor improvement in the performance of the models for describing Gaia data, the overall fits to the spectral energy distributions of individual stars were not improved over those using opacity of \citet{KS06}. Most importantly, the Gaia mass problem persists. With the negative result, \citet{STK25} speculate that the $\rm H^-$ bound-free opacity may be overestimated by a factor of 5 and point to uncertainty in this absorption mechanism as a potential source of the Gaia mass bias. We notice that this would be rather surprising, as the $\rm H^-$ bound-free absorption is relatively straightforward to model, in comparison to Ly $\alpha$ opacity, and its computation has been a topic of early "classical" and very accurate computational studies (e.g., \citet{C45,CE58,DFE66}). Nevertheless, the analysis of \citet{STK25} shows that none of these ad hoc corrections could consistently reconcile models and data at Hertzsprung-Russell diagrams constructed with photometric bandpasses of different wavelengths. 

In order to understand the true origin of the discrepancy between models and Gaia data, we undertook a detailed analysis of the models above and below $T_{\rm eff}\rm =6000\,K$, with the aim to identify significant differences and correlate these with the Gaia mass problem. In particular, we consider chemical equilibrium, a key factor not analysed in up to date studies.

\begin{figure}[t!]
   \centering
   \includegraphics[width=\hsize]{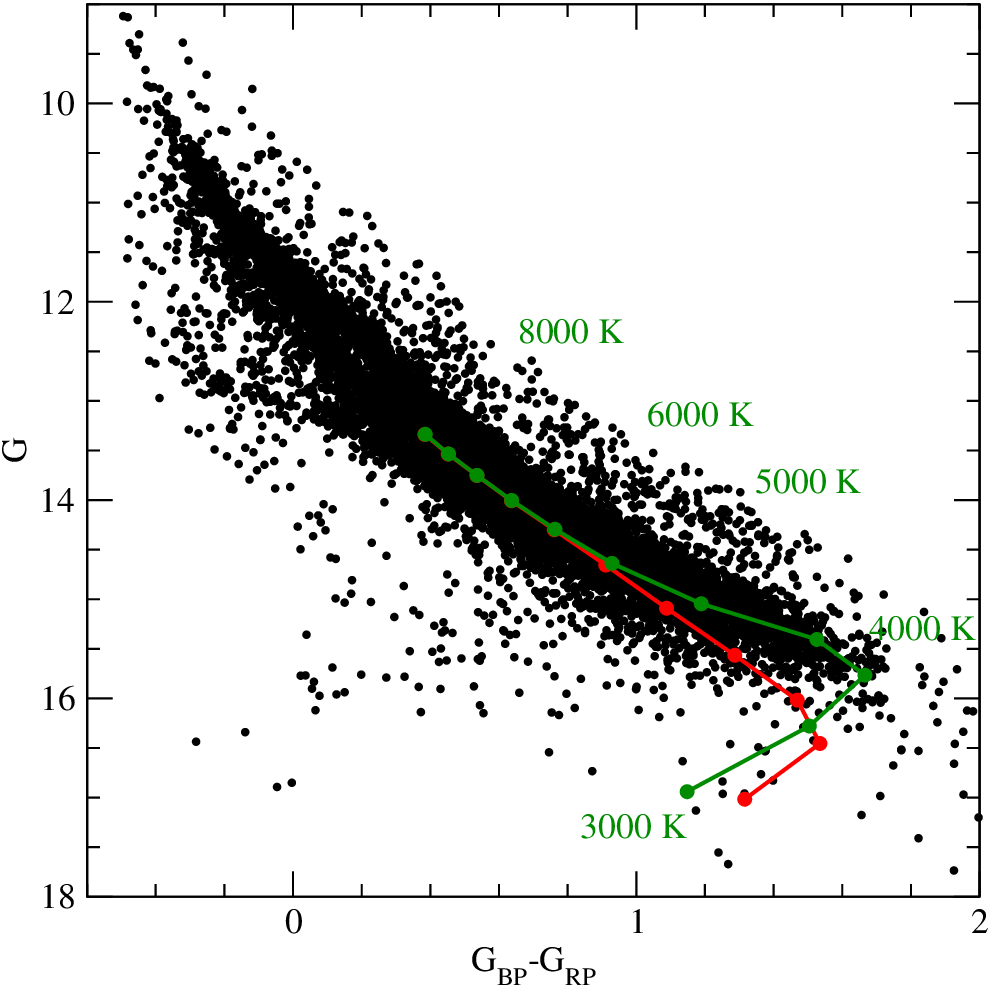}
      \caption{The Hertzsprung-Russell (H-R) diagram for the 100 pc sample of white dwarfs: Gaia DR3 data (filled black dots, \citet{JTR22}), pure hydrogen atmosphere cooling sequence of $log\,g=8$ with (red dots and solid line) and without (green dots and solid line) $\rm H_3^+$ species considered in the chemical equilibrium. Dots represent the effective temperature sequence from $8000\rm\,\rm K$ to $3000\rm\,\rm K$, from top to bottom, with an interval of $500\,\rm K$.}
         \label{F1}
   \end{figure}

\section{Computational approach}
We used the state-of-the-art white dwarf atmosphere models that include essential absorption mechanisms, $\rm Ly$ $\alpha$ red wing opacity \citep{KS06} and negative hydrogen ion bound-free absorption \citep{J88}. The models assume the presence of the following hydrogen species: $\rm H$, $\rm H^+$, $\rm H_2$, $\rm H_2^+$, $\rm H^-$, 
$\rm H_3^+$ and free electrons. These models have been successfully applied in reproduction of entire spectral energy distributions of several cool white dwarfs with hydrogen atmosphere (e.g., \citet{KS06,KKM08,SHK14}). Because atmospheres of stars at the very cool end become condensed, we account for the refractive effects in the radiative transfer equation \citep{KS04}. The refractive index has been estimated using Clausius–Mossotti relation \citep{R32} with polarizabilities of hydrogen atoms and hydrogen molecules of $3.8\cdot10^{-25}\,\rm cm^{2}$ and $8.2\cdot10^{-25}\,\rm cm^{2}$, respectively.

The quantum mechanical calculations were performed with the CPMD code (www.cpmd.org) using density functional theory with the PBE \citep{PBE96} approximation of the exchange-correlation functional. 

The Gaia sample of white dwarfs within $100\,\rm pc$ used in the analysis is that of \citet{JTR22}.

\section{Results and Discussion}
In Figure 1 we compare the cooling track of a hydrogen atmosphere white dwarf of surface gravity $\log g=8\,(cgs)$ (corresponding to an average white dwarf of mass $M\approx0.6\,M_{\sun}$) with the Gaia data. The deviation between the model and data for $T_{\rm eff}<6000\,\rm K$ is clearly visible. The discrepancy between Gaia and synthetic magnitudes results in incorrect assignment of lower masses to stars at the end of the cooling track \citep{BTK23,STK25}. 

\begin{figure*}[t!]
   \centering
   \includegraphics[width=0.95\hsize]{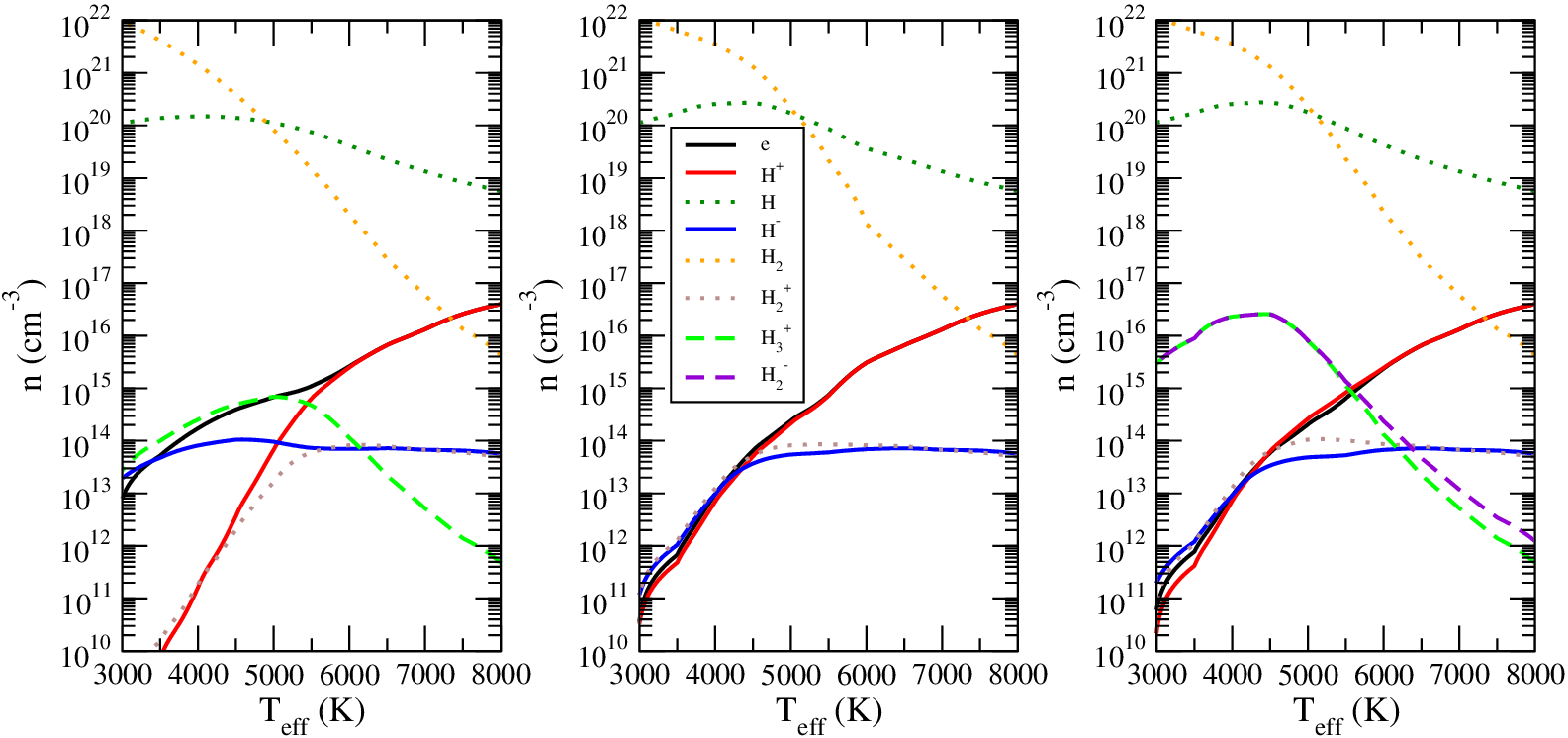}
      \caption{The photospheric (at the Rosseland mean optical depth, $\tau_R=2/3$) abundance of species  in the atmosphere of hydrogen white dwarf of $log\,g=8$ as a function of the effective temperature. Different colors and lines represent different species as indicated in the legend insert. The panels show results with (left) and without (middle) $\rm H_3^+$ species considered. The right panel shows the result obtained by assuming formation of a hypothetical anionic species (represented by $\rm H_2^-$ with the decreased formation energy).}
         \label{F2}
   \end{figure*}

In order to understand the origin of the discrepancy between the models and Gaia data, we check the variation in the atmospheric abundance of different hydrogen species with the effective temperature. As illustrated in Fig. \ref{F2}, for stars with $T_{\rm eff}>6000\,\rm K$, the dominant species is atomic hydrogen, which is the main donor of free electrons that determine the abundance of the negative hydrogen ion. For $T_{\rm eff}<6000\,\rm K$, molecular hydrogen forms in large quantities, contributing to the collision induced absorption (CIA) in the infrared \citep{SBT22}. Interestingly, together with the rise in the abundance of molecular hydrogen, the ionization equilibrium becomes affected by the formation of $\rm H_3^+$ species. These, having the lowest formation energy among the considered cations (Tab. \ref{T1}), become the most abundant positively charged species, significantly altering the abundance of free electrons and $\rm H^-$ species. 

Interestingly, the deviation between the models and the Gaia data is clearly correlated with the formation of $\rm H_3^+$ species. In order to check the impact of $\rm H_3^+$ formation on synthetic spectra and Gaia magnitudes, we computed models with suppressed abundance of this species. The resulting chemical equilibrium at $T_{\rm eff}<6000\,\rm K$ becomes drastically different. It is now governed by the ionization of $\rm H$ atoms, with some influence of $\rm H_2^+$ species, leading to a significant decrease in the abundance of free electrons and $\rm H^-$. The resulting cooling sequence overlaps with the Gaia data (Fig. \ref{F1}; and with other data: Fig. \ref{FS1}), strongly suggesting that the formation of $\rm H_3^+$ species is the main cause of mismatch between the models and the data. Suppression of $\rm H_3^+$ formation leads to significant decrease in the strength of $\rm H^-$ bound-free opacity, which depends on the number of absorbing $\rm H^-$ species. We notice that this decrease grows monotonically with lowering $T_{\rm eff}$, producing a gradual effect starting below $6000\,\rm K$. The overall effect is qualitatively different from the case of applying a constant change to the strength of the absorption \citep{STK25}. Although the simple modification of the chemical equilibrium by forbidding formation of $\rm H_3^+$ species already reconciles the models and the data, reasoning for it requires better understanding.

It is the ionization energy of species such as $\rm H$, $\rm H_2$ and formation energy of $\rm H_3^+$ that determine the ionization equilibrium and the resulting number of free electrons and $\rm H^-$. In Table \ref{T1} we provide the known, measured values of formation energies of these species. As mentioned, the $\rm H_3^+$ species has the lowest formation energy among cations and, once formed, has to significantly contribute to the ionization equilibrium. It is also a known hydrogen species detected in astronomical objects \citep{MTG20} with well constrained formation energy \citep{RPM04,RPS05}. Simple exclusion of its formation is thus not a plausible solution. 

Compared to the other hydrogen species, description of thermodynamic parameters of $\rm H_3^+$ is a challenge. Its partition function estimates, to which the abundance of $\rm H_3^+$ species is directly proportional, for $T>1000\rm \,K$ vary by at least an order of magnitude between different studies   \citep{NT95,RT11,CGP91,MSM10}, and in current models the upper values of \citet{NT95} are used. 
One plausible solution to the overabundance of $\rm H_3^+$ species is thus overestimation of its partition function. Indeed, the partition function of \citet{RT11} obtained with the advanced path integral Monte Carlo simulations, being by factor of $\sim 12$ lower, results in much lower abundance of $\rm H_3^+$ and $\rm H^-$, and better match of the models to the Gaia data (Figs. \ref{FS2} and \ref{FS3}).

Assuming that $\rm H_3^+$ forms as the chemical equilibrium predicts, there is a possibility that it is another, unaccounted for anionic species that compensates the excess positive charge caused by the appearance of $\rm H_3^+$, preventing formation of additional electrons, and indirectly $\rm H^-$. One natural candidate would be $\rm H_2^-$, which may form with the decrease in the effective temperature due to the increased number of molecular species. The existence of this species has been reported in some studies \citep{JKG11}, although as a metastable species, with the measured and computed here formation energy provided in Tab. \ref{T1}. We note that the measured value indicates that it should be still $\rm H^-$ to form as a counter charge to the $\rm H_3^+$ formation. On the other hand, the computed formation energy is significantly smaller (also bearing in mind the uncertainty of DFT) and just slightly larger than the energy of $\rm H_2$, which makes plausible that this anion species is more stable than measured. In Fig. \ref{F2} we demonstrate that when we decrease the formation energy further by $2.4\,\rm eV$, such an anion becomes dominant, making $\rm H_3^+$ ineffective in forming free electrons. The resulting ionization equilibrium, atmosphere models and the cooling track should be indistinguishable from the ones produced by suppressing $\rm H_3^+$ formation. The same effect is expected when another hydrogen-based anion forms, e.g., $\rm H_3^-$, which our calculations suggest may be indeed preferred to form over a pair of $\rm H_2$ + $\rm H^-$, although by only $0.06\rm \,eV$ (Tab. \ref{T1}). 
We note that the formation energy of $\rm H_3^-$ is not reported in the main databases \citep{RPM04,RPS05} but existing computational data suggest that it is chemically bounded \citep{BTD06}. Because of an already complex hydrogen chemical equilibrium at $T_{\rm eff}<6000\,\rm K$ shown by the models, there is also a possibility that more complex and exotic hydrogen species form, affecting the chemical and ionization equilibrium \citep{TM19}. Quantifying these would require accurate quantum mechanical calculations of a series of such possible species as well as derivation of the thermodynamic parameters of $\rm H_3^-$, which is a challenging task, as the case of $\rm H_3^+$ shows. We hope this study will trigger such an effort.

\begin{table}[h!]
\caption{The computed and measured \citep{RPM04,RPS05} formation energy (enthalpy) of different hydrogen species. The values are given in eV. The energies vs. formation of hydrogen atom (hydrogen molecule) are present. \label{T1}}
\centering
\begin{tabular}{lll}
\hline\hline
Species & Computed & Measured\\
\hline
$\rm H$    & 0 (2.27)    & 0 (2.24)\\
$\rm H^+$ &13.59 (15.86) & 13.60 (15.84)\\
$\rm H^-$ &-0.76 (1.50) & -0.75 (1.48)\\
$\rm H_2$     &  -4.53 (0) & -4.48 (0)\\
$\rm H_2^+$      &  10.60 (15.13) & 10.95 (15.42)\\
$\rm H_2^-$      & -4.51 (0.025) & -1.85 (2.63)\\
$\rm H_3$     & -4.36 (2.44) & -\\
$\rm H_3^+$      & 4.45 (11.25) & 4.79 (11.51)\\
$\rm H_3^-$      & -5.36 (1.44) & - \\

\hline
\end{tabular}
\end{table}

\begin{figure*}[t!]
   \centering
   \includegraphics[width=0.77\hsize]{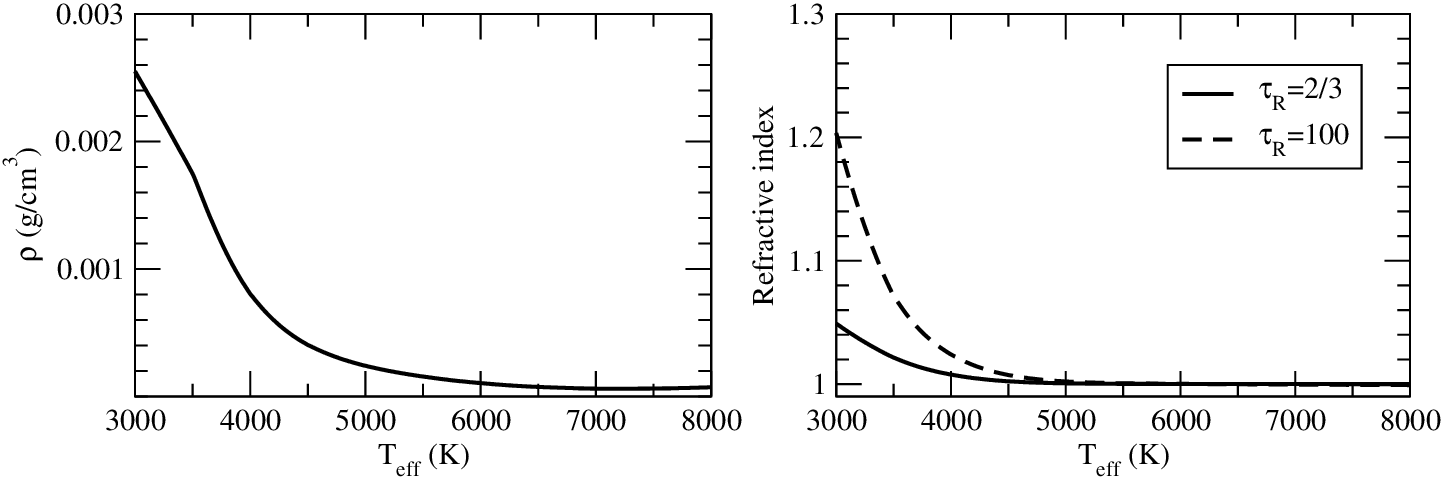}
      \caption{The photospheric (at the Rosseland mean optical depth, $\tau_R=2/3$) density (left panel) and refractive index at $\tau_R=2/3$ and $100$ (right panel), as a function of the effective temperature.}
         \label{F3}
   \end{figure*}

\citet{STK25} speculate that it is the non-ideal effects in a dense atmosphere that contribute to a potential distortion of the $\rm H^-$ opacity. To check if such a scenario is likely, in Fig. \ref{F3} we plotted the photospheric density profiles and refractive index as a function of the effective temperature. Indeed, although the density is still rather low ($<<0.1\,\rm g/cm^3$), the non-ideal effects may play a role, but at the very cool end ($T_{\rm eff}<<5000\,\rm K$), which is illustrated by an increase of the refractive index. The previous estimates show that the ionization energy of $\rm H^-$, as a descriptor of perturbation in a dense medium, could be affected, but only when densities exceed $0.1\,\rm g/cm^3$ \citep{K10}, much higher values than those encountered in cool, hydrogen atmosphere white dwarfs (Fig. \ref{F3}).

Another intriguing possibility is that the ionization equilibrium, namely number of free electrons, is somehow determined by the most abundant species only, hydrogen atom and molecule, through direct ionization. This will exclude formation of trace species, such as $\rm H_3^+$, from affecting the abundance of free electrons and $\rm H^-$. We note that this speculative assumption would indicate a non equilibrium process taking place. However, if present for any reason, it may also affect chemical equilibrium and ionization fraction in more extreme and challenging to model, helium-rich atmospheres \citep{KMS07}.

Nevertheless, we note that the cooling track obtained by removing $\rm H_3^+$ from chemical equilibrium not only follows the Gaia data well, but also indicates the abrupt of the cooling track at the effective temperature of just below $4000\,\rm K$. This is consistent with previous estimates of the effective temperature of white dwarfs at the end of the cooling sequence observed in the Galactic Disk (e.g., \citet{K07}).

\section{Conclusions}
We found here that the Gaia mass problem for cool white dwarfs shows a strong correlation with the dramatic change in the chemical/ionization equilibrium in hydrogen atmosphere of these stars. The white dwarf atmosphere models can be reconciled with Gaia data when the impact of $\rm H_3^+$ ion on the ionization equilibrium is suppressed and the number of free electrons is governed by the most abundant species, namely hydrogen atoms. Assuming expected formation of $\rm H_3^+$ species, 
this may indicate overestimation of the partition function of $\rm H_3^+$ and its abundance, a missing anionic species unaccounted for by the models (for instance $\rm H_3^-$), or that ionization equilibrium is somehow determined by the dominant species. Independently of which mechanism is realised, Gaia data call for the reconsideration of models of chemical equilibrium applied in the modelling of cool white dwarfs atmospheres.

\begin{acknowledgements}
The author gratefully acknowledges the computing time granted on the supercomputer JURECA at Forschungszentrum Jülich (Project cjiek61).
\end{acknowledgements}

\bibliographystyle{aa}
\bibliography{m}

\begin{appendix}

\section{Supplementary magnitude-color diagrams.}

As the supplementary information we present a set of complementary figures. These were created using data from Montreal White Dwarf Database 
(www.montrealwhitedwarfdatabase.org, \citet{DBC17}).

\begin{figure*}[]
    \centering
    \includegraphics[width=1.\textwidth]{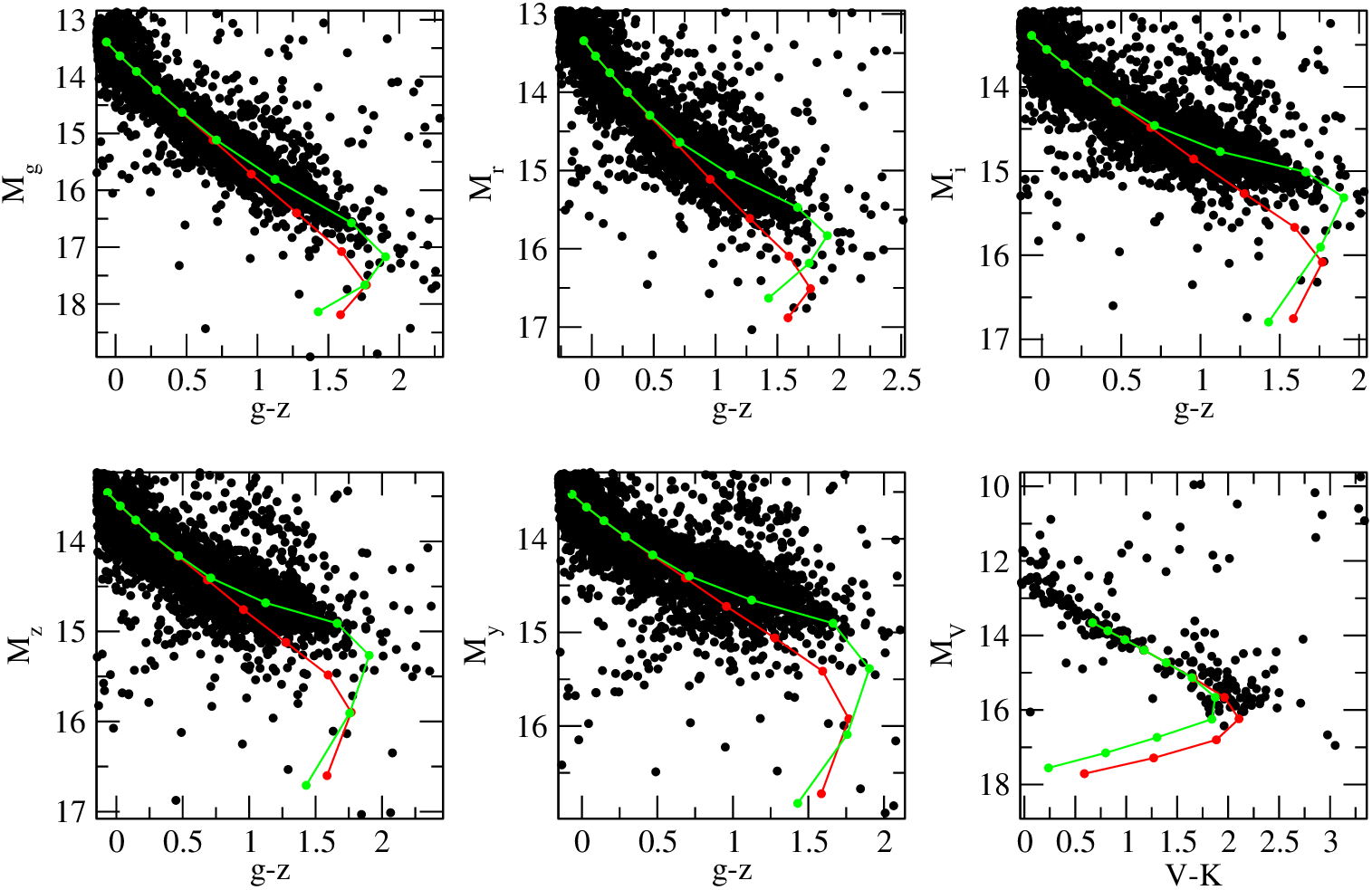}
\caption{The Hertzsprung-Russell (H-R) diagram for the sample of white dwarfs with PAN-STARRS nad BVK photometry (filled black dots, \citet{DBC17}), pure hydrogen atmosphere cooling sequence of $log\,g=8$ with (red dots and solid line) and without (green dots and solid line) $\rm H_3^+$ species considered in the chemical equilibrium. Dots represent the effective temperature sequence from $8000\rm\,\rm K$ to $3000\rm\,\rm K$, from top to bottom, with an interval of $500\,\rm K$. \label{FS1}}
\end{figure*}

In Fig. \ref{FS1} we present series of magnitude-color diagrams produced with PAN-STARRS and BVK photometry data. This figure is supplementary to Fig. \ref{F1} presented in the main text. 

\section{The Hertzsprung-Russell (H-R) diagram with the alternative partition function of $\rm H_3^+$.}

In Fig. \ref{FS2} we present the reproduction of Fig. \ref{F1} with the additional cooling sequence modelled with the $\rm H_3^+$ partition function of \citet{RT11}. We note that this sequence results in the improved match to the Gaia data, especially for $T_{\rm eff}>5000\,\rm K$. 

\begin{figure}[]
    \includegraphics[width=0.4\textwidth]{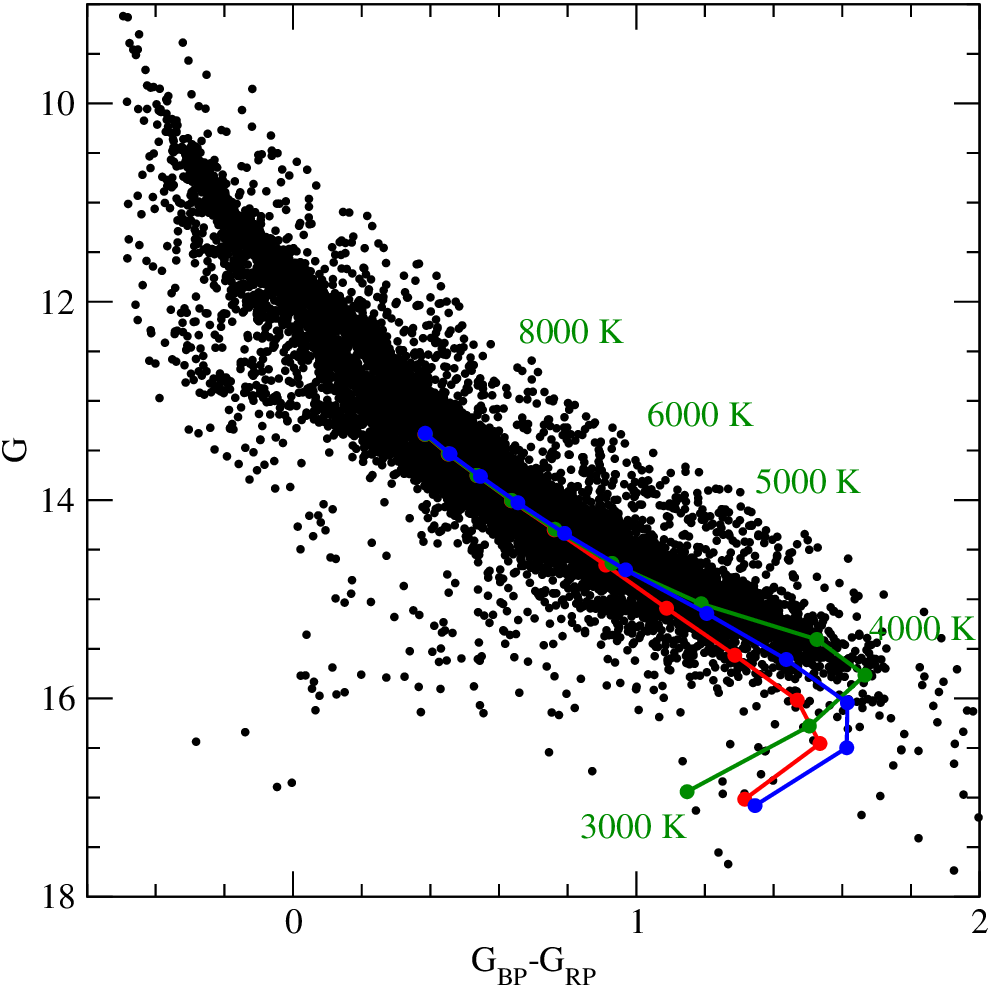}
\caption{The Hertzsprung-Russell (H-R) diagram. Data and models as in Fig. \ref{F1}.\label{FS2}. The blue line and filled dots represent the cooling sequence computed assuming $\rm H_3^+$ partition function of \citet{RT11}. \label{FS2}}
\end{figure}

\section{The abundances of species obtained with the alternative partition function of $\rm H_3^+$.}

\begin{figure}[]
    \includegraphics[width=0.3\textwidth]{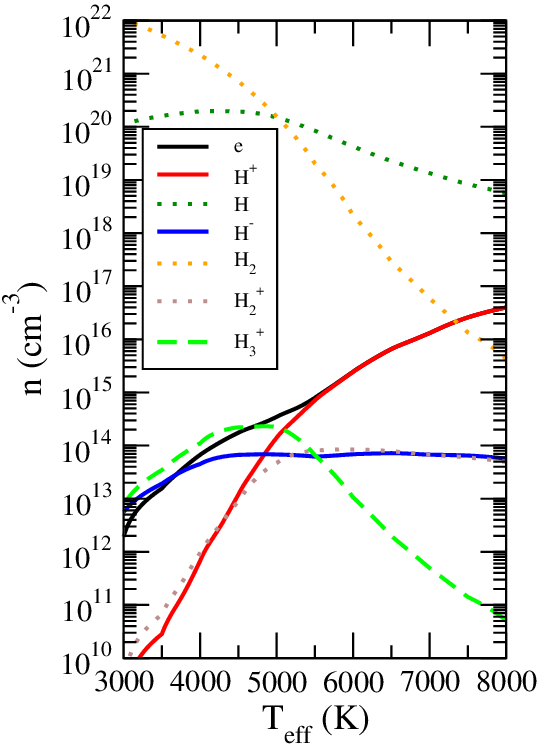}
\caption{The photospheric (at the Rosseland mean optical depth, $\tau_R=2/3$) abundance of species in the atmosphere of hydrogen white dwarf of $log\,g=8$ as a function of the effective temperature, obtained assuming the $\rm H_3^+$ partition function of \citet{RT11}. \label{FS3}}
\end{figure}

In Fig. \ref{FS3} we provide the atmospheric abundances of species computed assuming the $\rm H_3^+$ partition function of \citet{RT11}.

\section{Pure-H atmosphere models.}

\begin{figure*}[]
    \centering
    \includegraphics[width=0.8\textwidth]{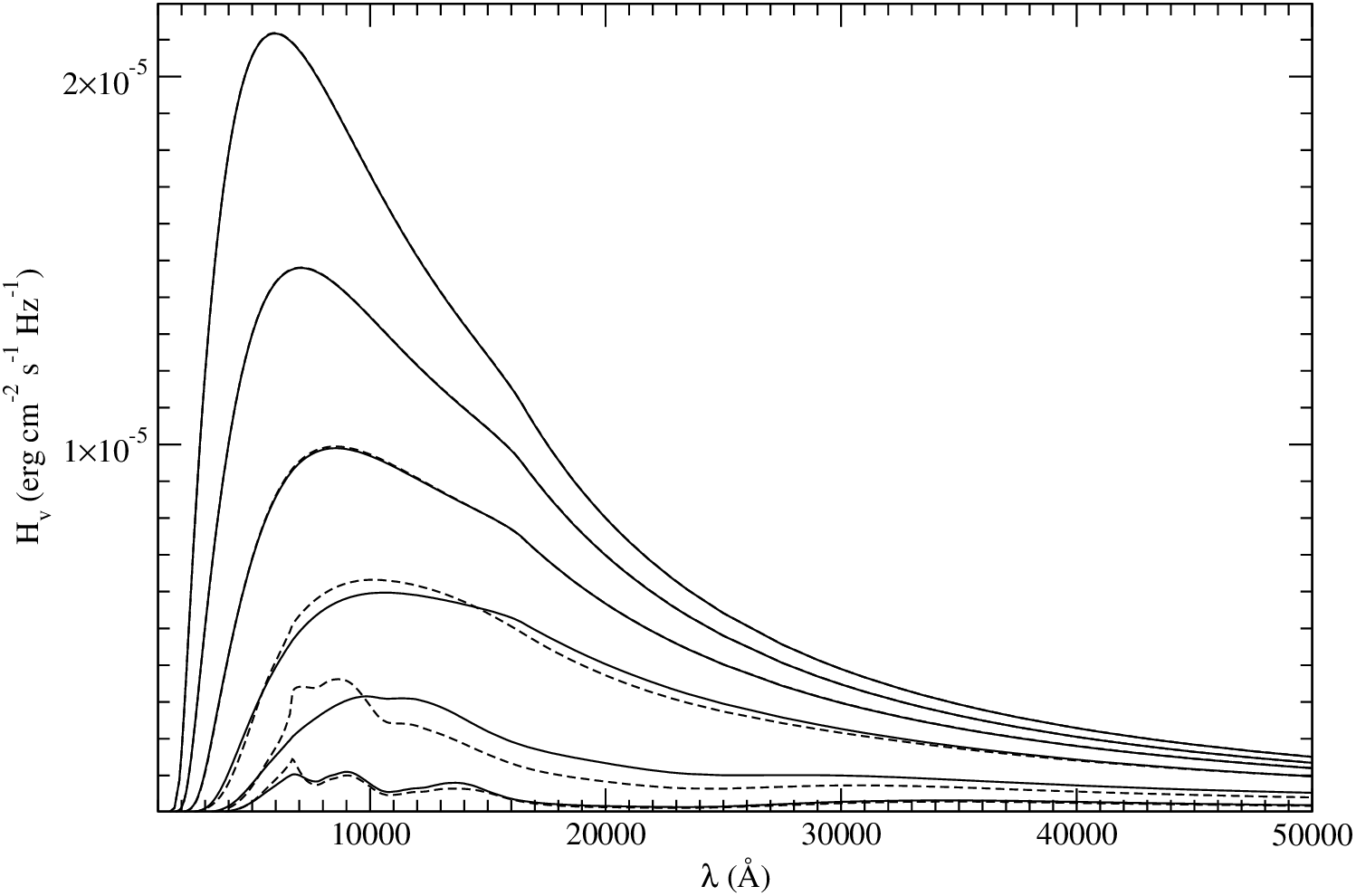}
\caption{The sequence of synthetic pure-H atmosphere models with (solid lines) and without (dashed lines) $\rm H_3^+$ species considered. The models represent the effective temperature sequence from $8000\rm\,\rm K$ to $3000\rm\,\rm K$, from top to bottom, with an interval of $1000\,\rm K$. Weak hydrogen lines present in the hotter atmospheres are removed for better visibility of spectral energy distributions.\label{FS4}}
\end{figure*}

In Fig. \ref{FS4} we present series of synthetic, pure-H atmosphere models computed with and without $\rm H_3^+$ species. The models show influence of $\rm H_3^+$ suppression for $T_{\rm eff}<6000\rm \, K$.

\end{appendix}

\end{document}